\documentclass[draftclsnofoot,onecolumn,12pt]{IEEEtran}
\usepackage{multirow}
\usepackage{booktabs}
\usepackage{lipsum}
\usepackage{amsfonts}
\usepackage{amsmath,cases,amssymb}
\usepackage{graphicx}
\usepackage{cite}
\usepackage{subfigure}
\usepackage{epsfig}
\usepackage{url}
\usepackage{algorithm}
\usepackage{algorithmic}
\usepackage{epstopdf}
\usepackage{balance}
\usepackage{color}
\usepackage[justification=centering]{caption}

\newcommand{\ds}{\displaystyle}
\newtheorem{myth}{Theorem}

\newcommand{\la}{\langle}
\newcommand{\ra}{\rangle}

\newcommand{\clL}{{\cal L}}
\newcommand{\clM}{{\cal M}}

\newcommand{\clP}{{\cal P}}

\newcommand{\clA}{{\cal A}}
\newcommand{\clB}{{\cal B}}
\newcommand{\clD}{{\cal D}}

\newcommand{\bt}{\mathbf{t}}
\begin{document}
\title{MIMO Beamforming for  Secure and Energy-Efficient Wireless Communication
\thanks{This work was supported in part  by the Australian Research Council’s Discovery Projects under Project DP130104617, in part by the U.K. Royal Academy of Engineering Research Fellowship under Grant RF1415$\slash$14$\slash$22, and in part by
the U.S. National Science Foundation under Grants CMMI-1435778 and ECCS-1647198. }}
\author{Nguyen T. Nghia$^1$, Hoang D. Tuan$^1$\thanks{$^1$Faculty of Engineering and Information Technology,
University of Technology, Sydney, NSW 2007, Australia (email:
nghia.t.nguyen@student.uts.edu.au, tuan.hoang@uts.edu.au).}, Trung Q. Duong$^2$\thanks{%
$^2$The School of Electronics, Electrical Engineering and Computer Science, Queen's University Belfast, Belfast BT7 1NN, United Kingdom (e-mail: trung.q.duong@qub.ac.uk).} and H. Vincent Poor$^3$\thanks{$^3$Department of Electrical Engineering, Princeton University, Princeton, NJ 08544 USA (e-mail: poor@princeton.edu).} }
\maketitle
\vspace*{-1cm}
\begin{abstract}
Considering a multiple-user multiple-input multiple-output (MIMO)  channel with an eavesdropper,
this letter develops a  beamformer design to optimize the energy efficiency in terms of secrecy bits per Joule under
secrecy  quality-of-service constraints. This is a very difficult design problem with no available exact solution techniques. A path-following procedure, which iteratively improves its feasible points by using a simple quadratic program of moderate
dimension, is proposed. Under any fixed computational tolerance the procedure terminates after finitely many iterations, yielding at least  a locally optimal solution.  Simulation results show the superior performance of the obtained algorithm over other existing methods.
\end{abstract}

\vspace*{-0.3cm}
\section{Introduction}  Secure communication achieved by exploiting the wireless physical layer
to provide secrecy in data transmission, has drawn  significant recent research attention (see e.g. \cite{Koyl11,P12,Hetal15} and references therein). The performance of this type of secure communication can be measured in terms of the
secrecy throughput, which  is the capacity
of conveying  information to the intended users while keeping it confidential from eavesdroppers \cite{P12,Cetal13}.
On the other hand, energy efficiency (EE)  has emerged as another important figure-of-merit in assessing the
performance of communication systems \cite{Caval14,Ietal14}. For most systems, both security and energy efficiency are of interest, and thus it is of interest to combine these two metrics into a single performance index called
the secrecy EE (SEE), which can be expressed in terms of secrecy bits per Joule.

Transmit beamforming can be used to enhance the two conflicting targets for optimizing SEE in multiple-user multiple-input
multiple-output (MU-MIMO) communications: mitigating MU interference  to maximize the users' information throughput,
and jamming eavesdroppers to control the leakage of information. However, the current approach to treat both EE \cite{TTJ15,VTFH16} and SEE \cite{ZYS15,TKT16}  is based on
costly zero-forcing beamformers, which completely cancel the MU interference and  signals received at the eavesdroppers.
The EE/SEE objective is in the form of a ratio of a concave function and a convex function, which can
be optimized by using  Dinkelbach's algorithm \cite{Dinkelbach1967}.
Each Dinkelbach's iteration still requires a log-det function
optimization, which is  convex but computationally quite complex. Moreover, zero-forcing beamformers are mostly suitable for low code rate applications and are applicable to specific MIMO systems only. {\color{black}The computational complexity of
SEE for single-user MIMO/SISO communications as considered in \cite{Kaetal15} and \cite{ZLJ16} is also high as each iteration still involves a difficult
nonconvex optimization problem.}

This letter aims to design transmit beamformers to optimize SEE subject to per-user secrecy  quality-of-service (QoS) and transmit power constraints. The specific contributions are detailed in the following dot-points.
\begin{itemize}
\item A path-following computational procedure, which invokes a simple convex quadratic program at each iteration and converges to at least a locally optimal solution, is proposed. The MU interference and eavesdropped  signals are effectively suppressed for optimizing the SEE. In contrast to  zero-forcing beamformers, higher code rates not only result in transmitting more concurrent data streams but also lead to much better SEE performance in our proposed beamformer design.
\item As a by-product, other important problems in secure and energy-efficient communications, such as EE maximization
subject to the secrecy level or
sum secrecy throughput maximization, which are still quite open for research, can be effectively addressed by the proposed procedure.
\end{itemize}
\textit{Notation.} All variables are written in boldface. For illustrative purpose, $f(\mathbf{V})$ is a mapping
of variable $\mathbf{V}$ while $f(\bar{V})$ is the output of mapping $f$ corresponding to a particular input $\bar{V}$.
$I_n$ denotes the identity matrix of size $n\times n$. The notation $(\cdot)^H$ stands for the Hermitian transpose, $|A|$ denotes the determinant of a square matrix $A$, and $\la A\ra$ denotes its trace
while $(A)^2 = AA^H$.
 The inner product $\langle{X,Y}\rangle$ is defined as $\la X^H Y\ra$ and therefore the Frobenius squared norm of a matrix $X$ is $||X||^2=\la XX^H\ra$. The notation $A\succeq B$ ($A\succ B$, respectively) means that $A-B$ is a positive semidefinite (definite, respectively) matrix. $\mathbb{E}[\cdot]$ denotes  expectation  and $\Re\{\cdot\}$ denotes the real part of a complex number. ${\cal CN}(0,a)$ denotes a circularly-symmetric complex Gaussian random variable with mean zero and variance $a$.
\vspace*{-0.3cm}
\section{SEE Formulation}
Consider a MIMO  system consisting of $D$ transmitters and $D$ users indexed by $1,\dots,D$.
Each transmitter $j$ is  equipped with $N$ antennas to transmit  information
to its intended user $j$ equipped with $N_r$ antennas. There is an eavesdropper equipped with $N_e$ antennas, which is
part of the legitimate network \cite{Koyl11,Cetal13}. The channel matrices
$H_{\ell j}\in\mathbb{C}^{N_r\times{N}}$ and $H_{\ell e}\in\mathbb{C}^{N_e\times N}$
from  transmitter $\ell$ to user $j$ and to the eavesdropper, respectively, are  known at the transmitters
by using the channel reciprocity, feedback and learning mechanisms \cite{Koyl11,Cetal13,Detal10,LPW11}.

A complex-valued vector $s_{j}\in\mathbb{C}^{d_1}$ contains the information transmitter $j$  intends to convey to user $j$, where $\mathbb{E}\left[s_{j}s_{j}^H\right]=I_{d_1}$, and $d_1\leq
N$ is the number of concurrent data streams.
Denote by $\mathbf{V}_{j}\in\mathbb{C}^{{N}\times d_1}$ the complex-valued beamformer matrix for user $j$. The ratio $d_1/N$ is called the code rate of $\mathbf{V}_{j}$. For notational convenience, define $\clD \triangleq \{1,\dots,D\}$ and
$\mathbf{V} \triangleq [\mathbf{V}_{j}]_{j\in\clD}$.

The received signal at user $j$ and  the  signal received at the eavesdropper are
\begin{eqnarray}
y_{j}&=&H_{jj}\mathbf{V}_{j}s_{j}
+\ds\sum_{\ell\in\clD\setminus \{j\}}
H_{\ell j}\mathbf{V}_{\ell} s_{\ell}
 + \tilde{n}_{j},\label{fd1}\\
y_e&=&\ds\sum_{j=1}^DH_{je}\mathbf{V}_js_j+\tilde{n}_e, \label{d1}
\end{eqnarray}
where $\tilde{n}_{j}\in {\cal CN}(0,\sigma_j^2)$ and $\tilde{n}_e\in {\cal CN}(0,\sigma_e^2)$ are  additive noises.

By (\ref{fd1}), the rate of information $f_{j}$ leaked from user $j$ (in nats) is
\begin{equation}
        f_{j}(\mathbf{V})=  \ln\left|I_{N_r}+(\clL_j(\mathbf{V}_j))^2
        (\Psi_{j}(\mathbf{V})+\sigma_j^2I_{N_r})^{-1}\right|,             \label{fd3}
\end{equation}
where $\clL_j(\mathbf{V}_j)\triangleq H_{jj}\mathbf{V}_{j}$ and
$\Psi_{j}(\mathbf{V})\triangleq \sum_{\ell\in\clD\setminus\{j\}}
(H_{\ell j}\mathbf{V}_{\ell})^2$.\\
On the other hand, the wiretapped throughput for user $j$ at the eavesdropper is
\begin{equation}
f_{j,e}(\mathbf{V}) \triangleq \ln\left|I_{{N}_e} + (\clL_{j,e}(\mathbf{V}_j))^2(\Psi_{j,e}(\mathbf{V})+\sigma_e^2I_{{N}_e})^{-1}\right|, \label{fd4}
\end{equation}
where $\clL_{j,e}(\mathbf{V}_j)\triangleq H_{je}\mathbf{V}_{j}$ and
$\Psi_{j,e}(\mathbf{V}) \triangleq \sum_{\ell\in\clD\setminus\{j\}}(H_{\ell e}\mathbf{V}_{\ell})^2$.
The secrecy throughput in transmitting information $s_j$ to user $j$ while keeping it confidential
from the eavesdropper is defined as \cite{P12,Cetal13}
\begin{equation}\label{d2}
f_{j,s}(\mathbf{V})\triangleq f_j(\mathbf{V})-f_{j,e}(\mathbf{V}).
\end{equation}
Following \cite{Xiong2012}, the consumed power for signal transmission is modelled by
$P^{\rm tot}(\mathbf{V}) \triangleq \zeta P^t(\mathbf{V}) + P_c$,
where $P^t(\mathbf{V})\triangleq \sum_{j=1}^D ||\mathbf{V}_{j}||^2$ is the total transmit power of
 the transmitters and $\zeta$ and $P_c$ are the reciprocal of the drain efficiency of the power amplifier
 and  the circuit power, respectively.

Consider the following secure beamformer design to optimize the system's energy efficiency:
\begin{subequations}\label{fd5}
\begin{eqnarray}
\ds\max_{\mathbf{V}}\ \ds\frac{1}{P^{\rm tot}(\mathbf{V})}\sum_{j=1}^D(f_{j}(\mathbf{V})-f_{j,e}(\mathbf{V}))
 & \mbox{s.t.}& \label{fd5a} \\
|| \mathbf{V}_{j}||^2 \le P_{\max},\ j\in\clD,&&
\label{fd5b} \\
f_{j}(\mathbf{V})-f_{j,e}(\mathbf{V}) \ge r_{j}, j\in\clD,&&\label{fd5c}
\end{eqnarray}
\end{subequations}
where the constraints (\ref{fd5b}) limit the transmit power, while
(\ref{fd5c}) are the secrecy  QoS constraints.

It can be seen from their definitions (\ref{fd3}) and (\ref{fd4}) that both throughput  $f_j$ and
wiretapped  throughput  $f_{j,e}$ are very complicated functions of
the beamformer variable $\mathbf{V}$. The approach of \cite{TTJ15} and \cite{VTFH16} (to
EE) and  \cite{ZYS15} and \cite{TKT16} (to SEE) seeks $\mathbf{V}$ in the class of zero-forcing
beamformers $\Psi_j(\mathbf{V})\equiv 0$,
$j\in\clD$ and $\sum_{\ell\in\clD}(H_{\ell e}\mathbf{V}_{\ell})^2\equiv 0$ to cancel completely all the MU interference and
wiretapped signals. Each throughput $f_j$ becomes a log-det function of only $\mathbf{V}_j$.  Dinkelbach's algorithm is
then applied to compute a zero-forcing solution of (\ref{fd5}), which requires a log-det function optimization for each
iteration. Such optimization  is still computationally difficult
with no available polynomial-time solvers. Note that
the feasibility of the zero-forcing constraints imposes $N\geq N_e+d_1$ and $D(N+N_r-N_e-2d_1)\geq (D-1)d_1$ \cite{TKT16}.
Thus, there is not much freedom for optimizing zero-forcing beamformers whenever $N$ is not large.

In the next section, we will provide a completely new computational approach  to (\ref{fd5}) by effectively enhancing
its difficult  objective  and  constraints.
\vspace*{-0.3cm}
\section{Path-following computational procedure}
By introducing a variable $\bt$ satisfying the convex quadratic constraint
\begin{equation}\label{e1}
\zeta\ds\sum_{j=1}^D|| \mathbf{V}_{j}||^2+P^{\rm BS}\leq \bt,
\end{equation}
the optimization problem (\ref{fd5}) can be equivalently expressed as
\begin{equation}\label{fd5.1}
\ds\max_{\mathbf{V},\bt}\ \clP(\mathbf{V},\bt)\triangleq
\frac{1}{\bt}\sum_{j=1}^D(f_{j}(\mathbf{V})-f_{j,e}(\mathbf{V}))\ \mbox{s.t.}\ (\ref{fd5b}), (\ref{fd5c}).
\end{equation}
In what follows, a function $h$ is said to be a {\it minorant} ({\it majorant}, resp.)
of a function $f$ at a point $\bar{x}$ in
the definition domain $\mbox{dom}(f)$ of $f$ iff $h(\bar{x})=f(\bar{x})$ and $h(\mathbf{x})\leq f(\mathbf{x})$ $\forall\ \mathbf{x}\in\mbox{dom}(f)$  ($h(\mathbf{x})\geq f(\mathbf{x})$ $\forall\ \mathbf{x}\in\mbox{dom}(f)$, resp.) \cite{Tuybook}.

By \cite{TTN16}, a {\it concave quadratic minorant} of the throughput function $f_j(\mathbf{V})$ at $V^{(\kappa)}\triangleq[V^{(\kappa)}_{j}]_{j\in\clD}$, which is feasible for (\ref{fd5b})-(\ref{fd5c}) is
\begin{equation}\label{thetajk}
\Theta^{(\kappa)}_{j}(\mathbf{V}) \triangleq a_{j}^{(\kappa)}+2\Re\{\clA_j^{(\kappa)},\clL_{j}(\mathbf{V}_{j})\ra\}-\la \clB_{j}^{(\kappa)}, \clM_{j}(\mathbf{V})\ra,
\end{equation}
where $\clM_{j}(\mathbf{V})\triangleq \Psi_{j}(\mathbf{V})+(\clL_{j}(\mathbf{V}_{j}))^2$,
$0>a_{j}^{(\kappa)}\triangleq f_{j}(V^{(\kappa)})-\la
(\clL_{j}(V^{(\kappa)}_{j}))^H(\Psi_{j}(V^{(\kappa)})+\sigma_j^2I_{N_r} )^{-1}\\ \clL_{j}(V^{(\kappa)}_{j})\ra
-\sigma_j^2\la (\Psi_{j}(V^{(\kappa)})+\sigma_j^2I_{{N}_r})^{-1}
-(\clM_{j}(V^{(\kappa)})+\sigma_j^2I_{{N}_r})^{-1}\ra$,
$\clA_j^{(\kappa)}\triangleq (\Psi_{j}(V^{(\kappa)})+\sigma_j^2I_{{N}_r})^{-1}\clL_{j}(V^{(\kappa)}_{j})$ and
\[
0\preceq\clB_j^{(\kappa)}\triangleq (\Psi_{j}(V^{(\kappa)})+\sigma_j^2I_{{N}_r})^{-1}
-(\clM_{j}(V^{(\kappa)})+\sigma_j^2I_{{N}_r})^{-1}.
\]
To provide a minorant of the secrecy throughput $f_{j,s}$ (see (\ref{d2})) at $V^{(\kappa)}$,
the next step is to find a {\it majorant} of the eavesdropper  throughput  function $f_{j,e}(\mathbf{V})$ at $V^{(\kappa)}$.
Reexpressing $f_{j,e}$ by
\begin{equation}
\ln\left|I_{{N}_e}+\clM_{j,e}(\mathbf{V})/\sigma_e^2\right|
-\ln\left|I_{{N}_e}+\Psi_{j,e}(\mathbf{V})/\sigma_e^2\right|,\label{fd4.1}
\end{equation}
for $\clM_{j,e}(\mathbf{V})\triangleq \Psi_{j,e}(\mathbf{V})+(\clL_{j,e}(\mathbf{V}_{j}))^2$,
and applying Theorem \ref{baseth} in the appendix for  upper bounding the first term  and lower bounding
the second term in (\ref{fd4.1}) yields the following {\it convex quadratic majorant} of $f_{j,e}$ at $V^{(\kappa)}$:
\[
\begin{array}{lll}
\Theta_{j,e}^{(\kappa)}(\mathbf{V})&\triangleq& a_{j,e}^{(\kappa)}-2\ds\sum_{\ell\in\clD\setminus\{j\}}\Re\{\la
H_{\ell e}V^{(\kappa)}_{\ell}\mathbf{V}_{\ell}^HH_{\ell e}^H \ra\}/\sigma_e^2\\
&&+\ds\la \clB^{(\kappa)}_{j,e1},\clM_{j,e}(\mathbf{V})\ra/\sigma_e^2+\la \clB^{(\kappa)}_{j,e2},\Psi_{j,e}(\mathbf{V})\ra/\sigma_e^2,
\end{array}
\]
where $a_{j,e}^{(\kappa)}\triangleq f_{j,e}(V^{(\kappa)})+\la (I_{N_e}+\clM_{j,e}(V^{(\kappa)})/\sigma_e^2)^{-1}-I_{N_e}
+\Psi_{j,e}(V^{(\kappa)})/\sigma_e^2\ra$, and
\[
\begin{array}{c}
0\preceq \clB^{(\kappa)}_{j,e1}\triangleq (I_{N_e}+\clM_{j,e}(V^{(\kappa)})/\sigma_e^2)^{-1},\\
0\preceq \clB^{(\kappa)}_{j,e2}\triangleq (\sigma_e^2)^{-1}I_{N_e}-(\sigma_e^2I_{N_e}+\Psi_{j,e}(V^{(\kappa)}))^{-1}.
\end{array}
\]
A {\it concave quadratic minorant}  of the secrecy throughput function $f_{j,s}$ at $V^{(\kappa)}$ is then
\begin{eqnarray}
\Theta^{(\kappa)}_{j,s}(\mathbf{V})&=&\Theta^{(\kappa)}_j(\mathbf{V})-\Theta^{(\kappa)}_{j,e}(\mathbf{V})\nonumber\\
&=&a^{(\kappa)}_{j,s}+\clA^{(\kappa)}_{j,s}(\mathbf{V})-\clB^{(\kappa)}_{j,s}(\mathbf{V}).\label{sratel}
\end{eqnarray}
Here, $a^{(\kappa)}_{j,s} \triangleq a^{(\kappa)}_{j}+a^{(\kappa)}_{j,e}$,
$\clA^{(\kappa)}_{j,s}(\mathbf{V})\triangleq 2\Re\{\la\clA_j^{(\kappa)},\clL_{j}(\mathbf{V}_{j})\ra\}
+2\sum_{\ell\in\clD\setminus\{j\}}\Re\{\la
H_{\ell e}V^{(\kappa)}_{\ell}\mathbf{V}_{\ell}^HH_{\ell e}^H \ra\}/\sigma_e^2$, and
$\clB^{(\kappa)}_{j,s}(\mathbf{V})\triangleq \la \clB_{j}^{(\kappa)}, \clM_{j}(\mathbf{V})\ra+
\ds\la \clB^{(\kappa)}_{j,e1},\clM_{j,e}(\mathbf{V})\ra+\la \clB^{(\kappa)}_{j,e2},\Psi_{j,e}(\mathbf{V})\ra/\sigma_e^2$.

Therefore, the nonconvex secrecy QoS constraints (\ref{fd5c}) can be innerly approximated by the following convex quadratic constraints {\color{black}in the sense that the feasibility of the former is guaranteed by the feasibility of the latter}:
\begin{equation}\label{kappa1bc}
\ds\Theta^{(\kappa)}_{j,s}(\mathbf{V})
\geq r_{j}, j=1,...,D.
\end{equation}
For good approximation, the following trust region is imposed:
\begin{eqnarray}
\clA^{(\kappa)}_{j,s}(\mathbf{V})&\geq& 0, \ j=1,...,D.\label{co2a}
\end{eqnarray}
By using the inequality
\[
\frac{x}{t}\geq 2\frac{\sqrt{x^{(\kappa)}}\sqrt{x}}{t^{(\kappa)}}-\frac{x^{(\kappa)}}{(t^{(\kappa)})^2}t\quad\forall
x>0, x^{(\kappa)}>0, t>0, t^{(\kappa)}>0
\]
we obtain $\clA^{(\kappa)}_{j,s}(\mathbf{V})/\bt
\geq \varphi_{j,s}^{(\kappa)}(\mathbf{V},\bt)$, for
\begin{equation}\label{e4a}
\varphi_{j,s}^{(\kappa)}(\mathbf{V},\bt)\triangleq
2b_{j,s}^{(\kappa)}\sqrt{\clA^{(\kappa)}_{j,s}(\mathbf{V})}-
c_{j,s}^{(\kappa)}\bt
\end{equation}
where $0<b_{j,s}^{(\kappa)}\triangleq \ds\sqrt{\clA^{(\kappa)}_{j,s}(V^{(\kappa)})}/t^{(\kappa)}$,
$0<c_{j,s}^{(\kappa)}\triangleq (b_{j,s}^{(\kappa)}/t^{(\kappa)})^2$, which is a
concave function \cite{Tuybook}.

With regard to $a_{j,s}^{(\kappa)}/\bt$  we define a concave function $a_{j,s}^{(\kappa)}(\bt)$ as follows:
\begin{itemize}
\item  If $a_{j,s}^{(\kappa)}<0$, define $a_{j,s}^{(\kappa)}(\bt)\triangleq a_{j,s}^{(\kappa)}/\bt$, which  is a
 concave function;
\item If $a_{j,s}^{(\kappa)}>0$, define $a_{j,s}^{(\kappa)}(\bt)=a_{j,s}^{(\kappa)}(2/t^{(\kappa)}-\bt/(t^{(\kappa)})^2)$,
which is a linear minorant of the convex function $a_{j,s}^{(\kappa)}/\bt$ at $t^{(\kappa)}$.
\end{itemize}
A {\it concave minorant} of $\Theta^{(\kappa)}_{j,s}(\mathbf{V})/\bt$, which is also a minorant of
$(f_j(\mathbf{V})-f_{j,e}(\mathbf{V}))/\bt$ at $(V^{(\kappa)}, t^{(\kappa)})$, is thus
\begin{equation}\label{e6}
g^{(\kappa)}_{j,s}(\mathbf{V},\bt)\triangleq a^{(\kappa)}_{j,s}(\bt)+\varphi_{j}^{(\kappa)}(\mathbf{V}_{j},\bt)-
\clB_{j,s}^{(\kappa)}(\mathbf{V})\ra/\bt.
\end{equation}
We now solve the nonconvex optimization problem (\ref{fd5}) by generating the next feasible point
$(V^{(\kappa+1)}, t^{(\kappa)})$ as the optimal solution of the following convex quadratic program (QP), which is an inner approximation \cite{Tuybook} of the nonconvex optimization problem (\ref{fd5.1}):
\begin{eqnarray}
\ds\max_{\mathbf{V},\mathbf{t}}\ \clP^{(\kappa)}(\mathbf{V},\mathbf{t})\triangleq
\ds\sum_{j=1}^Dg^{(\kappa)}_{j,s}(\mathbf{V},\bt)\nonumber\\
\mbox{s.t.}\quad (\ref{fd5b}), (\ref{e1}), (\ref{kappa1bc}), (\ref{co2a}).\label{kappa1}
\end{eqnarray}
Note that (\ref{kappa1}) involves $n=2DNd_1+1$ scalar real variables and $m=2D+1$ quadratic constraints so its computational complexity is $\mathcal{O}(n^2m^{2.5}+m^{3.5})$.\\
It can be seen that
${\cal P}(V^{(\kappa+1)}, t^{(\kappa+1)})\geq {\cal P}^{(\kappa)}(V^{(\kappa+1)}, t^{(\kappa+1)})
>{\cal P}^{(\kappa)}(V^{(\kappa)}, t^{(\kappa)})={\cal P}(V^{(\kappa)}, t^{(\kappa)})$
as long as $(V^{(\kappa+1)}, t^{(\kappa+1)})\neq (V^{(\kappa)}, t^{(\kappa)})$, i.e.
$(V^{(\kappa+1)}, t^{(\kappa+1)})$  is  better than $(V^{(\kappa)}, t^{(\kappa)})$.
This means that, once initialized from a feasible point
$(V^{(0)},t^{(0)})$ for (\ref{fd5.1}), the $\kappa$-th QP iteration
(\ref{kappa1}) generates a sequence $\{(V^{(\kappa)}, t^{(\kappa)})\}$ of feasible
and improved points toward the nonconvex optimization problem (\ref{fd5.1}),
which converges at least to a locally optimal solution of
(\ref{fd5}) \cite{TTN16}. Under the stopping criterion
\[
\left|\left(\clP(V^{(\kappa+1)}, t^{(\kappa+1)})- \clP(V^{(\kappa)},t^{(\kappa)})\right)/\clP(V^{(\kappa)},t^{(\kappa)})\right| \leq \epsilon
\]
for a given tolerance $\epsilon>0$, the QP iterations will terminate after finitely many iterations.

\begin{algorithm}[!t]
\caption{Path-following Algorithm for SEE Optimization}\label{alg1}
\begin{algorithmic}
\STATE \textit{Initialization:} Set $\kappa:=0$, and choose a feasible point $(V^{(0)}, t^{(0)})$ for (\ref{fd5.1}).
\STATE \textit{$\kappa$-th iteration:}  Solve (\ref{kappa1}) for an optimal solution $(V^{*},t^*)$ and set $\kappa:=\kappa+1$, ${V}^{(\kappa)}, t^{(\kappa)})\triangleq ({V}^{*}, t^{*})$ and calculate
$\clP(V^{(\kappa)}, t^{(\kappa)})$. Stop if $\left|\left({{\cal P}({V}^{(\kappa)},t^{(\kappa)}) -
{\cal P}({V}^{(\kappa-1)})}, t^{(\kappa-1)}\right)/{{\cal
P}({V}^{(\kappa-1)}, t^{(\kappa-1)})}\right|$ $\leq \epsilon$.
\end{algorithmic}
\end{algorithm}

The proposed path-following procedure  for computational solution of the nonconvex optimization problem (\ref{fd5}) is summarized in Algorithm~\ref{alg1}.

We note that a feasible initial point $(V^{(0)}, t^{(0)})$ for (\ref{fd5.1}) can be found by solving
\[
\max_{\mathbf{V}}\min_{j\in\clD}\
(f_{j}(\mathbf{V})-f_{j,e}(\mathbf{V}))/r_{j}\quad \mbox{s.t.}\quad  (\ref{fd5b})
\]
by the iterations
$\left\{\ds\max_{\mathbf{V}}\min_{j\in\clD}\
\Theta^{(\kappa)}_{j,s}(\mathbf{V})/r_{j}\quad
\mbox{s.t.}\quad  (\ref{fd5b})\right\}$,  which terminate upon reaching
$
(f_{j}(V^{(\kappa)})-f_{j,e}(V^{(\kappa)}))/r_{j} \geq 1  \ \forall j\in\clD,
$ to satisfy (\ref{fd5b})-(\ref{fd5c}).

The following problem of EE optimization under users' throughput QoS constraints and secrecy levels:
\begin{eqnarray}
\ds\max_{\mathbf{V}}\ \ds\frac{1}{P^{\rm tot}(\mathbf{V})}\sum_{j=1}^Df_{j}(\mathbf{V})
\quad \mbox{s.t.}\quad (\ref{fd5b}),\nonumber\\
 f_{j}(\mathbf{V}) \ge r_{j}\ \& \ f_{j,e}(\mathbf{V})\leq \epsilon, j=1,...,D,\label{tst1}
\end{eqnarray}
where $\epsilon$ is set small enough to keep the users' information confidential from the eavesdropper, is simpler than
(\ref{fd5}). It can be addressed by a similar path-following procedure, which solves the following QP at the
$\kappa-$th iteration instead of (\ref{kappa1}):
\begin{subequations}\label{cee}
\begin{eqnarray}
\ds\max_{\mathbf{V},\bt}\sum_{j=1}^D\left(a_{j}^{(\kappa)}/\bt+
4b_j^{(\kappa)}\sqrt{\Re\{\la \clA_j^{(\kappa)},\clL_{j}(\mathbf{V}_{j})\ra\}}\right.\nonumber\\
\ds\left.-2c_j^{(\kappa)}\bt-\la \clB_{j}^{(\kappa)}, \clM_{j}(\mathbf{V})\ra/\bt\right)\quad\mbox{s.t.}\quad(\ref{fd5b}),\label{ceea}\\
\Re\{\la\clA_j^{(\kappa)},\clL_{j}(\mathbf{V}_{j})\ra\}\geq 0,\ j\in \clD,\label{ceeb}\\
\Theta^{(\kappa)}_j(\mathbf{V})\geq r_j\ \&\
\Theta^{(\kappa)}_{j,e}(\mathbf{V})\leq \epsilon, \ j\in\clD,\label{ceec}
\end{eqnarray}
\end{subequations}
where $0<b_j^{(\kappa)}\triangleq \la(\clL_{j}(V^{(\kappa)}_{j}))^H(\Psi_{j}(V^{(\kappa)})+\sigma_j^2I_{{N}_r})^{-1}\clL_{j}(V^{(\kappa)}_{j})\ra^{1/2}/t^{(\kappa)}$,
 $0<c_{j}^{(\kappa)}\triangleq (b_{j}^{(\kappa)}/t^{(\kappa)})^2$ and  $\clA_j^{(\kappa)}$ and
$\clB_{j}^{(\kappa)}$ are defined from (\ref{thetajk}). A feasible initial point  $(V^{(0)}, t^{(0)})$ for (\ref{tst1}) can be found by solving
\[
\max_{\mathbf{V}}\min_{j\in\clD}\min\{f_{j}(\mathbf{V})-r_j,\epsilon-f_{j,e}(\mathbf{V})\}\quad\mbox{s.t.}\quad  (\ref{fd5b})
\]
by the iterations
\[
\max_{\mathbf{V}}\min_{j\in\clD}\min\{
\Theta^{(\kappa)}_{j}(\mathbf{V})-r_{j}, \epsilon-\Theta^{(\kappa)}_{j,e}(\mathbf{V})\}\quad
\mbox{s.t.}\quad  (\ref{fd5b})\},
\]
which terminate upon reaching
$f_{j}(V^{(\kappa)})-r_j\geq 0$, $\epsilon-f_{j,e}(V^{(\kappa)})\geq 0$  $\forall j\in\clD$,
to satisfy (\ref{fd5b}), (\ref{tst1}).

Lastly, the problem of  sum secrecy throughput maximization
\[
\max_{\mathbf{V}}\ \sum_{j=1}^D(f_{j}(\mathbf{V})-f_j(\mathbf{V}))
 \quad \mbox{s.t.}\ (\ref{fd5b}), (\ref{fd5c})
\]
is also simpler than the SEE optimization problem (\ref{fd5}), which can be addressed by a similar
path-following procedure with the  QP
\[
\max_{\mathbf{V}}\
\sum_{j=1}^D\Theta^{(\kappa)}_{j,s}(\mathbf{V})\
\mbox{s.t.}\  (\ref{fd5b}),  (\ref{kappa1bc})
\] solved at the $\kappa-$th iteration instead of (\ref{kappa1}).
\vspace*{-0.4cm}
\section{Numerical examples}
The fixed parameters are: $D=3$, $N=12$, $N_r=6$, $N_e=9$, $\sigma_j\equiv 1$, $\sigma_e=1$,
$r_j\equiv 1$ bits/s/Hz, $\zeta=1$ and $P_c\in \{7, 10\}$ dB.
The secrecy level $\epsilon=0.05/\log_2e$ is set in solving (\ref{tst1}).
The channels are Rayleigh fading so
their coefficients are generated as ${\mathcal{CN}}(0,1)$.

\begin{figure}[h]
  \centering
   \vspace*{-0.2cm}
  \includegraphics[width=0.9\textwidth,]{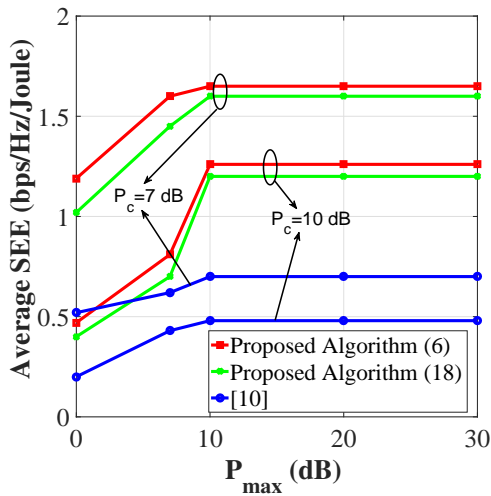}
  \caption{Average SEE vs. $P_{max}$ for $d_1=3$. }
  \label{fig1}
\end{figure}

\begin{figure}[h]
  \centering
  \includegraphics[width=0.9\textwidth,]{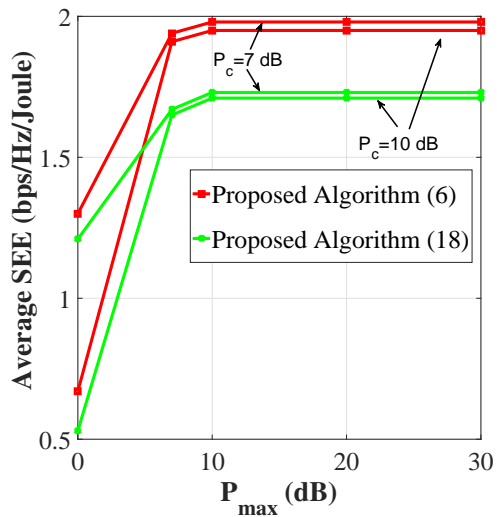}
  \caption{Average SEE vs. $P_{max}$ for $d_1=4$. }
  \label{fig2}
\end{figure}

For the first numerical example, the  number of data streams $d_1=3$ is set, so the code rate is $3/12=1/4$.
Each $\mathbf{V}_j$ is of size $12\times 3$. Figure \ref{fig1} shows the SEE performance of our proposed beamformer and
the zero-forcing beamformer  \cite{ZYS15,TKT16}. One can see that the former outperforms the
latter substantially. Apparently, the latter is not quite suitable for both EE and SEE.
The SEE performance achieved by the formulation (\ref{fd5}) is better than that achieved  by
 the formulation (\ref{tst1}) because the secrecy level is enhanced with the users's throughput in the former instead of being constrained  beforehand in the latter. When the transmit power $P_{\max}$ is small, the denominator of  the
SEE objective in (\ref{fd5}) and (\ref{tst1}) is dominated by the constant circuit power $P_c$.
As a result, the SEE is maximized by maximizing its numerator, which is  the system sum secrecy throughput.
On the other hand, the SEE objective is likely maximized by minimizing the transmitted power $P_{\max}$ in its denominator when the latter is dominated by $P_{\max}$. That is why the SEE saturates once $P_{\max}$ is beyond a threshold
 according to Figure \ref{fig1}.
We increase the number $d_1$ of data streams   to $4$ in the second numerical example. The code rate is thus $4/12=1/3$.
For this higher-code-rate case, the  zero-forcing beamformers \cite{ZYS15,TKT16} are infeasible. Comparing Figure \ref{fig1} and Figure \ref{fig2} reveals  that higher code-rate beamforming is also much better in terms of SEE because it leads to greater freedom in designing $\mathbf{V}_j$ of size $12\times 4$ for maximizing the SEE. In other words, the effect of code rate diversity on the SEE is observed.
\vspace*{-0.4cm}
\section{Conclusion}\label{sec:Conclusion}
We have proposed a path-following computational procedure for the beamformer design to maximize the energy efficiency
of a secure MU MIMO wireless communication system and have also showed its potential in solving other important optimization problems in secure
and energy-efficient communications. Simulation results have confirmed the superior performance of the proposed method over the exiting techniques.

{\bf Acknowledgement.} The authors thank Dr. H. H. Kha for providing the computational code from \cite{TKT16}.
\vspace*{-0.4cm}
\section*{Appendix}
\begin{myth}\label{baseth} For a given $\sigma>0$, consider a function
\[
f(\mathbf{X})=\ln|I_m+(\mathbf{X})^2/\sigma|
\]
in $\mathbf{X}\in\mathbb{C}^{m\times n}$. Then for any $\bar{X}\in\mathbb{C}^{m\times n}$, it is true that
\begin{equation}\label{b1}
h(\mathbf{X})\leq f(\mathbf{X})\leq g(\mathbf{X})
\end{equation}
with the {\it concave} quadratic function
\begin{eqnarray}
h(\mathbf{X})=a_l
+2\Re\{\la \bar{X}\mathbf{X}^H \ra\}/\sigma
-\ds\la \clB_l,(\mathbf{X})^2\ra/\sigma
\label{b2}
\end{eqnarray}
and the {\it convex} quadratic function
\begin{eqnarray}
g(\mathbf{X})=a_u+\ds\la \clB_u,(\mathbf{X})^2\ra/\sigma
\label{b3}
\end{eqnarray}
where $a_l\triangleq f(\bar{X})-\la (\bar{X})^2\ra/\sigma$, $0\preceq \clB_l\triangleq \sigma^{-1} I_m-(\sigma I_m+(\bar{X})^2)^{-1}$, and
$a_u\triangleq f(\bar{X})+\la (I_m+(\bar{X})^2/\sigma)^{-1}- I_m\ra$,
$0\prec \clB_u\triangleq (I_m+(\bar{X})^2/\sigma)^{-1}$. Both functions $h$ and $g$ agree with $f$ at $\bar{X}$.
\end{myth}
{\it Proof.} Due to  space limitations, we provide only a sketch of the proof. Rewrite
$f(\mathbf{X})=-\ln|I_m-(\mathbf{X})^2/((\mathbf{X})^2+\sigma I_m)^{-1}|$, which is convex as a function in
$((\mathbf{X})^2,(\mathbf{X})^2+\sigma I_m)$ \cite{TTN16}. Then $h(\mathbf{X})$ defined by (\ref{b2}) actually is the first
order approximation of this function at $((\bar{X})^2,(\bar{X})^2+\sigma I_m)$, which is its minorant at $((\bar{X})^2,(\bar{X})^2+\sigma I_m)$ \cite{Tuybook}, proving the first inequality in (\ref{b1}).\\
On the other hand, considering $f$ as a concave function in $(\mathbf{X})^2$, $g(\mathbf{X})$ defined by (\ref{b3})
is seen as its first order
approximation at $(\bar{X})^2$ and  thus is its majorant at  $(\bar{X})^2$ \cite{Tuybook}, proving
the second inequality in (\ref{b1}).

\end{document}